\documentclass[american]{elsarticle}
\usepackage[LGR,T1]{fontenc}
\usepackage[latin9]{inputenc}
\usepackage{geometry}
\usepackage{calc}
\usepackage{graphicx}

\makeatletter

\DeclareRobustCommand{\greektext}{%
  \fontencoding{LGR}\selectfont\def\encodingdefault{LGR}}
\DeclareRobustCommand{\textgreek}[1]{\leavevmode{\greektext #1}}
\ProvideTextCommand{\~}{LGR}[1]{\char126#1}

\usepackage{hyperref}
\usepackage{url}

\makeatother

\usepackage{babel}
\begin{document}

\begin{frontmatter}{}

\title{Machine learning for protein folding and dynamics }

\author{Frank Noé }

\address{Department of Mathematics and Computer Science, Freie Universität
Berlin, Arnimallee 6, 14195 Berlin, Germany}

\ead{frank.noe@fu-berlin.de}

\author{Gianni De Fabritiis}

\address{Computational Science Laboratory, Universitat Pompeu Fabra, Barcelona
Biomedical Research Park (PRBB), Doctor Aiguader 88, 08003 Barcelona,
Spain, and 
Institucio Catalana de Recerca i Estudis Avanats (ICREA), Passeig
Lluis Companys 23, Barcelona 08010, Spain}

\ead{gianni.defabritiis@upf.edu }

\author{Cecilia Clementi }

\address{Center for Theoretical Biological Physics, and Department of Chemistry,
Rice University, 6100 Main Street, Houston, Texas 77005, United States}

\ead{cecilia@rice.edu}

\begin{abstract}
Many aspects of the study of protein folding and dynamics have been
affected by the recent advances in machine learning. Methods for the
prediction of protein structures from their sequences are now heavily
based on machine learning tools. The way simulations are performed
to explore the energy landscape of protein systems is also changing
as force-fields are started to be designed by means of machine learning
methods. These methods are also used to extract the essential information
from large simulation datasets and to enhance the sampling of rare
events such as folding/unfolding transitions. While significant challenges
still need to be tackled, we expect these methods to play an important
role on the study of protein folding and dynamics in the near future.
We discuss here the recent advances on all these fronts and the questions
that need to be addressed for machine learning approaches to become
mainstream in protein simulation.
\end{abstract}

\end{frontmatter}{}

\subsection*{Introduction}

During the last couple of decades advances in artificial intelligence
and machine learning have revolutionized many application areas such
as image recognition and language translation. The key of this success
has been the design of algorithms that can extract complex patterns
and highly non-trivial relationships from large amount of data and
abstract this information in the evaluation of new data. 

In the last few years these tools and ideas have also been applied
to, and in some cases revolutionized problems in fundamental sciences,
where the discovery of patterns and hidden relationships can lead
to the formulation of new general principles. In the case of protein
folding and dynamics, machine learning has been used for multiple
purposes \citep{Ma2015,Wang2017,ingraham2018learning,Xu465955,Wang2019,Noe_ARPC2019}.

As protein sequences contain all the necessary information to reach
the folded structure, it is natural to ask if the ideas and algorithms
that have proved very useful to associate labels to images can also
help to associate a folded structure to a protein sequence. Indeed,
protein structure prediction has greatly benefitted from the influx
of idea from machine learning, as it has been demonstrated in the
CASP competitions in the last few years, where several groups have
used machine learning approaches of different kinds \citep{Ma2015,Wang2017,Ovchinnikov294,ingraham2018learning},
and the AlphaFold team from DeepMind won the 2018 competition by a
margin \citep{AlphaFold2018,AlphaFold}.

In addition to protein structure prediction, machine learning methods
can help address other questions regarding protein dynamics. Physics-based
approaches to protein folding usually involve the design of an energy
function that guides the dynamics of the protein on its conformational
landscape from the unfolded to the folded state. Different ideas have
been used in the past several decades to design such energy functions,
from first-principle atomistic force field \citep{LindorffLarsen2012,Robustelli2018}
to simplified coarse-grained effective potential energies \citep{ClementiCOSB}
encoding physical principles such as for instance the energy landscape
theory of protein folding \citep{ClementiJMB2000,Davtyan2012}. In
this context, neural networks can help design these energy functions
to take into account of multi-body terms that are not easily modeled
analytically \citep{Wang2019}. 

Another aspect where machine learning has made a significant impact
is on the analysis of protein simulations. Even if we had an accurate
protein force-field and we could simulate the dynamics of a protein
long enough to sample its equilibrium distribution, there is still
the problem of extracting the essential information from the simulation,
and to relate it to experimental measurements. In this case, unsupervised
learning methods can help to extract metastable states from high dimensional
simulation data and to connect them to measurable observables \citep{NoeClementi_COSB17_SlowCVs}.

In the following we review the recent contributions of machine learning
in the advancement of these different aspects of the study of protein
folding and dynamics. As the field is rapidly evolving, most probably
these contributions will become even more significant in the near
future.

\subsection*{Machine learning for protein structure prediction}

Structure prediction consists in the inference of the folded structure
of a protein from the sequence information. The most recent successes
of machine learning for protein structure prediction arise with the
application of deep learning to evolutionary information \citep{marks2011protein,ovchinnikov2014robust}.
It has long been known that the mutation of one amino acid in a protein
usually requires the mutation of a contacting amino acid in order
to preserve the functional structure \citep{Altschuh1987,Gobel1994,Weigt_2008,Szurmant2018}
and that the co-evolution of mutations contains information on amino
acid distances in the three dimensional structure of the protein.
Initial methods \citep{marks2011protein,Morcos_2011} to extract this
information from co-evolution data were based on standard machine
learning approaches but later methods based on deep residual networks
have shown to perform better in inferring possible contact maps \citep{Ma2015,Wang2017}.
More recently, it has been shown that it is possible to predict a
distance matrices \citep{Xu465955} from co-evolutionary information
instead of just contact maps. This result was accomplished by using
a probabilistic neural network to predict inter-residue distance distributions.
From a complete distance matrix, it is relatively straightforward
to obtain a protein structure, but of course the prediction of the
distance matrix from co-evolution data is not perfect, nor complete.
Yet, in \citep{Ovchinnikov294} it was shown that, if at least 32-64
sequences are available for a protein family, then this data are sufficient
to obtain the fold class for 614 protein families with currently unknown
structures, when the co-evolutionary information is integrated in
the Rosetta structure prediction approach. Admittedly, the authors
concede that this is not yet equivalent to obtain the crystal structure
to the accuracy that would be useful, for instance, for drug discovery.
However, it still represents a major achievement in structure prediction.

Every two years, the performance of the different methods for structure
prediction is assessed in the CASP (Critical Assessment of Techniques
for Protein Structure Prediction) competition, where a set of sequences
with structures yet to be released are given to participants to predict
the structure blindly. The extent of the impact of machine learning
in structure prediction has been quite visible in the latest CASP
competitions. The typical methodology in previous CASP editions for
the top ranked predictions has been to use very complex workflows
based on protein threading and some method for structure optimization
like Rosetta \citep{Raman_2009}. Protein threading consists in selecting
parts of the sequence for which there are good templates in the PDB
and stitch them together \citep{Jones1992}. A force-field can then
then be used to relax this object into a protein structure. The introduction
of co-evolution information in the form of contact maps prediction
provided a boost in the performance, at the expense of even more complex
workflows. 

Historically the difference between top predictors in CASP has been
minimal \textendash{} indicating that there was not a clearly better
method, but rather an incremental improvement of the workflows. This
situation created a barrier of entry to a certain extent for new ideas
and models. However, in the latest edition of CASP (CASP13), the group
of AlphaFold \citep{AlphaFold} ranked first with a very simplified
workflow \citep{AlphaFold2018}, heavily based on machine learning
methods. The approach extended the contact and distance matrix predictions
to predict histograms of distances between amino acids using a very
deep residual network on co-evolutionary data. This approach allowed
to take into account implicitly the possible errors and inaccuracy
in the prediction itself. In addition, it used an autoencoder architecture
derived from previous work on drawing \citep{gregor2015draw} to replace
threading all-together and generate the structure directly from the
sequence and distance histograms. The use of an autoencoder guarantees
an implicit, but much more elegant threading of the available structural
information in the PDB to the predicted structure. In a second approach
from the same group, a knowledge-based potential derived from the
distance histograms was also used. The potential was simply minimized
to converged structures. This last protein-specific potential minimization
might look surprising at first, but it is actually very similar to
well known structure-based models for protein folding \citep{taketomi1975studies,ClementiJMB2000}. 

An alternative and interesting machine learning approach for structure
predictions, which also offers wider applicability, is to use end-to-end
differentiable models \citep{ALQURAISHI2019292,ingraham2018learning,Anand2018}.
While the performance of these methods does not yet reach the performance
of co-evolution based methods for cases where co-evolutionary information
is high, they can be applied to protein design, and in cases where
co-evolution data is missing. In \citep{ALQURAISHI2019292}, a single
end-to-end network is proposed that is composed by multiple transformations
from the sequence to the protein backbone angles and finally to three-dimensional
coordinates on which a loss function is computed in terms of root
mean square deviations against known structures. In \citep{ingraham2018learning}
a sequence-conditioned energy function is parameterized by a deep
neural network and Langevin dynamics is used to generate samples from
the distribution. In \citep{Anand2018} a generative adversarial model
is used to produce realistic $C_{\alpha}$ distance matrices on blocks
up to 128-residues, then standard methods are used to recreate the
backbone and side chain structure from there. Incidentally, a variational
autoencoder was also tested as a baseline with comparable results.
This model is not conditioned on sequence, so it is useful for generating
new structures and for in-painting missing parts in a crystal structure.

\subsection*{Folding proteins with machine learned force fields}

 State-of-the-art force fields can reproduce with reasonable accuracy
the thermodynamical and structural properties of globular proteins
\citep{LindorffLarsen2012} or intrinsically disordered proteins (IDPs)
\citep{Robustelli2018}. Generally, force fields are designed by first
assigning a functional form for all the different types of interactions
(e.g., electrostatic, Van der Waals, etc.) between the atoms of different
types, then optimizing the parameters in these interactions to reproduce
as best as possible some reference data. 

In the last few years, a new approach on the design of force-fields
has emerged, that takes advantage of machine learning tools \citep{BehlerParrinello_PRL07_NeuralNetwork,RuppEtAl_PRL12_QML}.
The idea is to use either a deep neural network or a some other machine
learning model to represent the classical energy function of a system
as a function of the atomic coordinates, instead of specifying a functional
form a priori \citep{Schuett2018}. The model can then be trained
on the available data to ``learn'' to reproduce some desired properties,
such as energies and forces as obtained from quantum mechanical calculations.
As a neural network is a universal function approximator, this approach
has the significant advantage that can approximate a large number
of possible functional forms for the energy, instead of being constrained
by a predefined one, and can in principle include multi-body correlations
that are generally ignored in classical force-fields. The downside
of this increased flexibility however resides in the fact that a very
large amount of data is needed to train the machine learning model
as the model may extrapolate poorly in regions of the conformational
space where data are not available. So far, large amount of quantum
chemical calculations have been used to train such force-fields, but
in principle experimental data could also be included \citep{Chen2018}.

The machine learning approach to force field design has evolved rapidly
in the last decade, but it has so far mostly been tested on small
organic molecules. Some of the proposed methods are tailored to reproduce
the thermodynamics of specific molecules (e.g., \citep{ChmielaEtAl_SciAdv17_EnergyConserving}),
while others attempt to design transferable force-field that are trained
on a large number of small molecules and could in principle be used
to simulate a much larger molecule such as a protein (e.g., \citep{Smith2017,Smith2018}).
Indeed, quantum mechanical calculations on water, amino acids, and
small peptides have been included in the latest generation of machine-learned
classical force-fields (e.g. the development version of the ANI potential
\citep{ANIgithub}). We are aware of one instance where a machine-learned
force-field has been used to simulate a 50 ns molecular dynamics trajectory
of a cellulose-binding domain protein (1EXG) in its folded state.
Recently, a transferable machine-learned force-field has been tested
on polypeptides. However, machine-learned force-fields have not (yet)
been used for protein folding simulations, nor have they been used
to predict thermodynamic or kinetic properties. While we believe that
this will be possible and machine-learned force-fields will be widely
used in protein simulations in the near future, at the moment there
are still some significant challenges that need to be overcome towards
this goal \citep{Noe_ARPC2019}. 

One fundamental challenge resides on the modeling of long-range interactions.
If only quantum calculations on small molecules are used in the training
of force-fields, interactions on scales larger than these molecules
could easily be missed in the training. The locality of the machine-learned
force-fields could be insufficient to capture electrostatic interactions,
or long-range van der Waals interactions \citep{Hermann_2017}. This
problem could be addressed by separating the long-range effects in
the force-field. For instance, atomic partial charges could be learned
\citep{Nebgen2018} simultaneously to local energy terms and used
in electrostatic interactions that they could be added to the machine-learned
energy part to obtain a total energy that is used in the training.

Another main challenge resides in the software used for the simulations.
Calculating energies and forces for a protein configuration by means
of a trained neural network is several orders of magnitude faster
than obtaining these quantities ab-initio with quantum mechanical
calculations, but it's still slower than with a standard classical
force-field. In order to simulate protein folding, molecular dynamics
trajectories of at least microseconds are needed and this timescale
is not currently accessible with machine-learned force-fields. Research
in this area has so far mostly focused on obtaining an accurate representation
for the energy and forces for molecules and tests have been performed
on small systems, mostly as a proof of concept. As this field mature,
we believe that significant efforts will also be made to optimize
the software for practical applications and molecular dynamics simulation
with machine learned force-fields will become a viable alternative
to current approaches. Additionally, the whole arsenal of methods
that have been developed to enhance the sampling of protein configurational
landscapes with classical force-fields (e.g., \citep{LaioParrinello_PNAS99_12562,PretoClementi_PCCP14_AdaptiveSampling})
can also be used with machine-learned force-fields to reach longer
timescales and larger system sizes. 

\begin{figure}

\centering{}\includegraphics[width=0.8\textwidth]{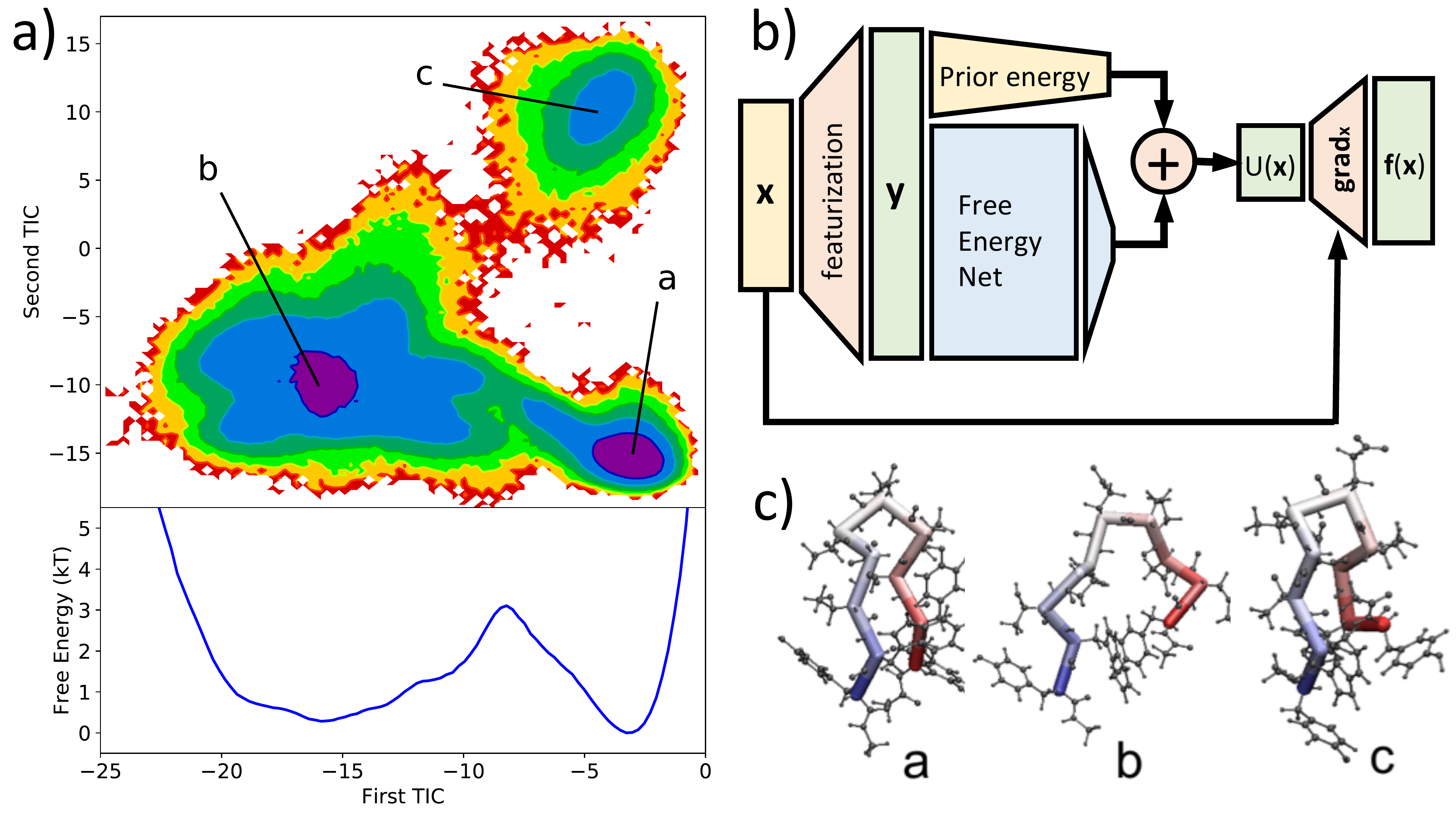}\caption{\label{fig-CGnet}(a) Folding free energy landscape of the protein
Chignolin as obtained with a coarse-grained model that uses a neural
network to represent the effective energy (CGnet). Top panel: Free
energy as obtained from CGnet, as a function of the first two collective
coordinates obtained with the Time-Lagged Indipendent Component Analysis
(TICA) method \citep{Perez-Hernandez2013}. Bottom panel: Projection
of the free energy on the first TICA coordinate. (b) The CGnet neural
network architecture. (c) Representative Chignolin configurations
in the three minima from (a). Figure adapted from \citep{Wang2019}.}
\end{figure}

\subsection*{Machine learning of coarse-grained protein folding models}

In parallel to efforts for the design of atomistic force-fields, machine
learning has also been used to obtain coarser models \citep{John2017,ZhangHan2018_CG,Wang2019},
that could be applied to study larger systems and longer timescales
with reduced computational resources. Coarse-grained models map groups
of atoms in some effective interactive ``beads'' and assign an effective
energy function between the beads to try to reproduce some properties
of a protein system. Different properties could be targeted, and different
strategies have been used to design coarse-grained models, either
starting from atomistic simulations (bottom-up) (e.g., \citep{Noid2008,Shell2008}),
experimental data (top-down) (e.g., \citep{Monticelli2008}) or enforcing
general ``rules'' such as the minimal frustration principle for
protein folding \citep{ClementiJMB2000,Davtyan2012}. In principle,
the same ideas used in the design of atomistic force-fields from quantum
mechanical data can be used to make the next step in resolution and
design coarse-grained molecular models from all-atom molecular simulations
\citep{ClementiCOSB}. One main problem in the design of models at
a resolution coarser than atomistic is the fact that by renormalizing
local degrees of freedom multi-body terms emerge in the effective
energy function even if only pairwise interactions were used in a
reference atomistic force-field. Such multi-body terms should then
be taken into account in the energy function of the coarse-grained
model to correctly reproduce the thermodynamics and dynamics of the
model at finer resolution. Attempts have been made to include these
terms in coarse-grained models, but it is challenging to define suitable
and general functional forms to capture these effect in an effective
energy function. For this reason, neural networks appear as a natural
choice for the design of coarse-grained potentials, as they can automatically
capture non-linearities and multi-body terms while agnostic on their
specific functional form. Indeed, in the last few years, several groups
have attempted to use machine learning methods to design coarse-grained
potentials for different systems \citep{John2017,ZhangHan2018_CG,Wang2019}.
Most recently, CGnet (see Figure \ref{fig-CGnet}), a neural network
for coarse-grained molecular force-fields, has been proposed and has
been used to model the folding/unfolding dynamics of a small protein
\citep{Wang2019}. The CGnet applications presented so far have been
system-specific. However similar ideas to what has been used in the
design of transferable atomistic force-fields from quantum mechanical
data could also been used to try to obtain more transferable coarse-grained
models. In general, transferability remains an outstanding issue in
the design of coarse models \citep{Noid2013} and its requirement
may decrease the ability of reproducing faithfully properties of specific
systems. So far, the challenges in the definition of general and multi-body
functional forms for coarse-grained models have not allowed to rigorously
investigate the trade-off between transferability and accuracy for
such models. The use of machine learning tools to design effective
potential energy functions may soon allow to explore this question
systematically.

\subsection*{Machine learning for analysis and enhanced simulation of protein
dynamics}

Machine learning has been quite impactful in the analysis of simulations
of protein dynamics. In this context, two closely related aims are:
1) the extraction of collective variables (CVs) associated with the
slowest dynamical processes and the metastable states (that can be
defined from the knowledge of the slow CVs) from given protein molecular
dynamics (MD) simulation data \citep{NoeClementi_COSB17_SlowCVs},
and 2) enhancing the simulations so as to increase the number of rare
event transitions between them.

A cornerstone for the extraction of slow CVs, metastable states and
their statistics are shallow machine learning methods such as Markov
state models (MSMs) \citep{PrinzEtAl_JCP10_MSM1} and Master-equation
models \citep{BucheteHummer_JPCB08}, which model the transitions
between metastable states via a Markovian transition or rate matrix.
A key advantage of MSMs is that they can be estimated from short MD
simulations started from an arbitrary (non-equilibrium) distribution,
and yet make predictions of the equilibrium distribution and long-timescale
kinetics. While more complex models, e.g. including memory, are conceivable,
MSMs are simpler to estimate, easier to interpret and are motivated
by the observation that if they are built in the slow CVs of the molecule,
the error made by the Markovian approximation is close to zero for
practical purposes \citep{PrinzEtAl_JCP10_MSM1}.

\begin{figure}
\centering{}\includegraphics[width=0.8\textwidth]{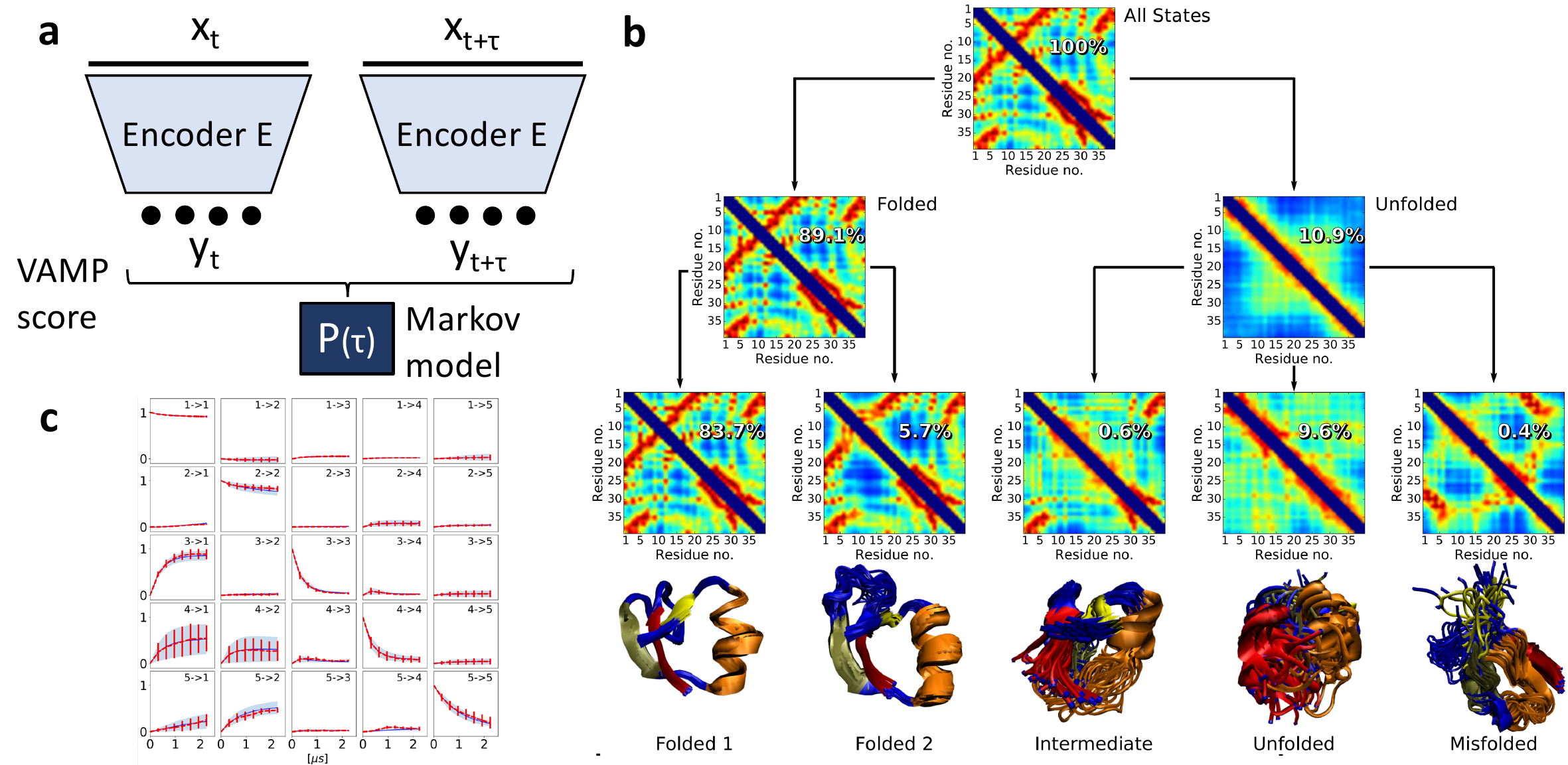}\caption{\label{fig:VAMPnet}\textbf{VAMPnet and application to NTL9 protein
folding}. \textbf{a}) a VAMPnet \citep{MardtEtAl_VAMPnets} includes
an encoder $E$ which transforms each molecular configuration $\mathbf{x}_{t}$
to a latent space of ``slow reaction coordinates'' $\mathbf{y}_{t}$,
and is trained on pairs $(\mathbf{y}_{t},\mathbf{y}_{t+\tau})$ sampled
from the MD simulation using the VAMP score \citep{WuNoe_VAMP}. \textbf{b})
Hierarchical decomposition of the NTL9 protein state space by a network
with two and five output nodes. Mean contact maps are shown for all
MD samples grouped by the network, along with the fraction of samples
in that group. 3D structures are shown for the five-state decomposition,
residues involved in \textgreek{a}-helices or \textgreek{b}-sheets
in the folded state are colored identically across the different states.
If the encoder performs a classification, the dynamical propagator
$\mathbf{P}(\tau)$ is a Markov state model.\textbf{ c}) Chapman\textendash Kolmogorov
test comparing long-time predictions of the Koopman model estimated
at $\tau=320$ ns and estimates at longer lag times. Figure modified
from \citep{MardtEtAl_VAMPnets}.}
\end{figure}

For this reason, much method development has been made in the past
10-15 years in order to optimize the pipeline for the construction
of MSMs, that is: finding suitable molecular features to work with
\citep{SchererEtAl_VAMPselection}, reducing the dimensionality of
feature space \citep{PerezEtAl_JCP13_TICA,SchwantesPande_JCTC13_TICA},
clustering the resulting space \citep{HusicPande_JCTC17_Ward,BucheteHummer_JPCB08},
estimating the MSM transition matrix \citep{TrendelkampSchroerEtAl_InPrep_revMSM}
and coarse-graining it \citep{DeuflhardWeber_LAA05_PCCA+,NoeEtAl_PMMHMM_JCP13}.
While all steps of this pipeline have significantly improved over
time, constructing MSMs this way is still very error prone and depends
on significant expert knowledge. A critical step forward was the advent
of the variational approach of conformation dynamics (VAC) \citep{NueskeEtAl_JCTC14_Variational}
and later the more general variational approach of Markov processes
(VAMP) \citep{WuNoe_VAMP}. These principles define loss functions
that the best approximation to the slow CVs should minimize, and can
thus be used to search over the space of features, discretization
and transition matrices variationally \citep{SchererEtAl_VAMPselection}.
Recently, VAMPnets have been proposed that use neural networks to
find the optimal slow CVs and few-state MSM transition matrices by
optimizing the VAMP score \citep{MardtEtAl_VAMPnets} (Fig. \ref{fig:VAMPnet}a),
and hence replace the entire human-built MSM pipeline by a single
end-to-end learning framework. VAMPnets have been demonstrated on
several benchmark problems including protein folding (Fig. \ref{fig:VAMPnet}b)
and have been shown to learn high-quality MSMs without significant
human intervention (Fig. \ref{fig:VAMPnet}c). When used with an output
layer that does perform a classification, VAMPnets can be trained
to approximate directly the spectral components of the Markov propagator
\citep{MardtEtAl_VAMPnets,CheeSidkyFerguson_arxiv19_ReversibleVAMPnets}.

The aim of enhancing MD sampling is closely connected to identifying
the metastable states or slow CVs of a given molecular system. As
the most severe sampling problems are due to the rare-event transitions
between the most long-lived states, such as folding/unfolding transitions,
identifying such states or the corresponding slow CVs on the fly can
help to speed up the sampling. So-called adaptive sampling methods
perform MD simulation in multiple rounds, and select the starting
states for the new round based on a model of the slow CVs or metastable
states found so far. Adaptive sampling for protein simulations has
been performed using MSMs \citep{DoerrDeFabritiis_JCTC14_OnTheFly,Hruska_2018}
and with neural network approximations of slow CVs \citep{Chen_2018,RibeiroTiwary_JCP18_RAVE}.
Since adaptive sampling uses unbiased (but short) MD trajectories
it is possible to reconstruct the equilibrium kinetics using MSMs,
VAMPnets or similar methods. Recently, adaptive sampling has been
used to sample protein-protein association and dissociation reversibly
in all-atom resolution, involving equilibrium timescales of hours
\citep{PlattnerEtAl_NatChem17_BarBar}.

An alternative to adaptive sampling is to use enhanced sampling methods
that speed up rare event sampling by introducing bias potentials,
higher temperatures, etc., such as umbrella sampling, replica-exchange
or metadynamics. Since these methods typically work in a space of
few collective variables, they are also sensitive to making poor choices
of collective variables, which can lead to sampling that is either
not enhanced, or even slower than the original dynamics. Machine learning
has an important role here as it can help these methods by learning
optimal choices of collective variables iteratively during sampling.
For example, shallow machine learning methods have been used to adapt
the CV space during Metadynamics \citep{McCartyParrinello_VACMetadynamics,SultanPande_JCTC17_TICAMetadynamics},
adversarial and deep learning have used to adapt the CV space during
variationally enhanced sampling (VES, \citep{ValssonParrinello_PRL14_VariationalMeta})
\citep{ZhangYanNoe_ChemRxiv19_TALOS,BonatiZhangParrinello_arxiv19_NeuralVES}.
A completely different approach to predict equilibrium properties
of a protein system is the Boltzmann Generator \citep{NoeEtAl_19_BoltzmannGenerators}
that trains a deep generative neural network to directly sample the
equilibrium distribution of a many-body system defined by an energy
function, without using MD simulation.

Since enhanced sampling changes the thermodynamic state of the simulation,
it is suitable for the reconstruction of the equilibrium distribution
at a target thermodynamic state by means of reweighting Boltzmann
probabilities, but generally loses information about the equilibrium
kinetics. Ways to recover the kinetics include: (i) extrapolating
to the equilibrium kinetics of rare event transitions by exploiting
the Arrhenius relation \citep{TiwaryParrinello_PRL14_MetadynamicsDynamics},
(ii) learning a model of the full kinetics and thermodynamics by combining
probability reweighting and MSM estimators in a multi-ensemble Markov
model \citep{WuEtAL_PNAS16_TRAM}, or (iii) reweighting transition
pathways \citep{DonatiKeller_JCP18_Girsanov}. Machine learning and
particularly deep learning has not been used much in these methods,
but certainly have potential to improve them.

\subsection*{Conclusions}

Machine learning can provide a new set of tools to advance the field
of molecular sciences, including protein folding and structure prediction.
Nonetheless, physical and chemical knowledge and intuition will remain
invaluable in the foreseeable future to design the methods and interpret
the results obtained. In particular, machine learning can help us
to extract new patterns from the data that are not immediately evident,
but in virtually all areas reviewed above, machine learning methods
that incorporate the relevant physical symmetries, invariances and
conservation laws perform better than black-box methods. Furthermore,
a trained scientist is still essential to provide meaning to the patterns
and use them to formulate general principles.

\textbf{Acknowledgement}

We gratefully acknowledge funding from European Research Council (ERC
CoG 772230 \textquotedblleft ScaleCell\textquotedblright{} to F.N.),
the Deutsche Forschungsgemeinschaft (CRC1114/A04 and GRK2433 DAEDALUS
to F.N.), the MATH+ Berlin Mathematics research center (AA1-6 and
EF1-2 to F.N.), the Einstein Foundation in Berlin (visiting fellowship
to C.C.), the National Science Foundation (grants CHE-1265929, CHE-1740990,
CHE-1900374, and PHY-1427654 to C.C. ), the Welch Foundation (grant
C-1570 to C.C.), MINECO (Unidad de Excelencia Mar\'{i}a de Maeztu
MDM-2014-0370 and BIO2017-82628-P to G.D.F), FEDER (to G.D.F), and
the European Union's Horizon 2020 research and innovation program
under grant agreement No 675451 (CompBioMed project to G.D.F).

\bibliographystyle{unsrt}

\end{document}